\documentclass[12pt,oneside,reqno]{amsart} 	
\usepackage[margin=1in]{geometry}			
\usepackage[parfill]{parskip}				
\usepackage{graphicx}
\usepackage{amssymb}
\usepackage{amsmath}					
\usepackage{epstopdf}
\usepackage[hidelinks]{hyperref} 
\title{A time-symmetric formulation of quantum entanglement}
\author{Michael B. Heaney\\3182 Stelling Drive\\Palo Alto, CA 94303\\mheaney@alum.mit.edu\\
Orcid ID: 0000-0002-6402-5923}
\date{30 January 2021}				
\begin{document}
\maketitle
\begin{abstract} 
I numerically simulate and compare the entanglement of two quanta using the conventional formulation of quantum mechanics and a time-symmetric formulation that has no collapse postulate. The experimental predictions of the two formulations are identical, but the entanglement predictions are significantly different. The time-symmetric formulation reveals an experimentally testable discrepancy in the original quantum analysis of the Hanbury Brown--Twiss experiment, suggests solutions to some parts of the nonlocality and measurement problems, fixes known time asymmetries in the conventional formulation, and answers Bell's question \textquotedblleft How do you convert an \textquoteleft and' into an \textquoteleft or'?"
\end{abstract}

\pagebreak
\section{Introduction} 
Smolin says ``the second great problem of contemporary physics [is to] resolve the problems in the foundations of quantum physics''~\cite{Smolin}. One of these problems is quantum entanglement, which is at the heart of both new quantum information technologies~\cite{IkeandMike} and old paradoxes in the foundations of quantum mechanics~\cite{Penrose}. Despite significant effort, a comprehensive understanding of quantum entanglement remains elusive~\cite{Horodecki}. In this paper, I compare how the entanglement of two quanta is explained by the conventional formulation of quantum mechanics~\cite{Commins,CTDL,Griffiths} and by a time-symmetric formulation that has no collapse postulate. The time-symmetric formulation and its numerical simulations can facilitate the development of new insights and physical intuition about entanglement. There is also always the hope that a different point of view will inspire new ideas for furthering our understanding of quantum behavior.

Time-symmetric explanations of quantum behavior predate the discovery of the Schr\"{o}dinger equation~\cite{Tetrode} and have been developed many times over the past century~\cite{Friederich}. {The time-symmetric formulation (TSF) used in this paper has been described in detail and compared to other TSF's before~\cite{HeaneyA,HeaneyB}. Note in particular that the TSF used in this paper is significantly different than the Two-State Vector Formalism (TSVF)~\cite{TSVF}. First, the TSVF postulates that a {particle} is completely described by a two-state vector, written as $\langle\phi\vert\thickspace\vert\psi\rangle$. This two-state vector is not mathematically defined. In contrast, the TSF postulates that the {transition} of a particle is completely described by a complex transition amplitude density $\phi^\ast\psi$, which is mathematically defined. The TSF defines this transition amplitude density as the algebraic product of the two wavefunctions, which is a dynamical function of position and time. Second, the TSVF postulates that wavefunctions collapse upon measurement~\cite{ACE1}, while the TSF postulates that wavefunctions never collapse.}

The particular time-symmetric formulation described in this paper is a type IIB model, in the classification system of Wharton and Argaman~\cite{Wharton}. It is called time-symmetric because (for symmetrical boundary conditions) the complex transition amplitude densities (defined below) are the same under a 180-degree rotation about the symmetry axes perpendicular to the time axes. The conventional formulation does not have this symmetry. To the best of my knowledge, this is the first quantitative explanation of entanglement by a time-symmetric formulation. The closest work appears to be~\cite{Sutherland,Sinha,Qureshi}. This paper extends the time-symmetric formulation described in~\cite{HeaneyA} from a single particle to two entangled particles. Preliminary results were presented at the 2019 Annual Meeting of the APS Far West Section.

Identical quanta have the same intrinsic physical properties, e.g., mass, electric charge, and spin. However, identical quanta are not necessarily indistinguishable: an electron in your finger and an electron in a rock on the moon are distinguishable by their location. Identical quanta can become indistinguishable when their wavefunctions overlap such that it becomes impossible, even in principle, to tell them apart.

Entanglement is usually taught using spin or polarization degrees of freedom. But entanglement also occurs in the spatial wavefunctions of systems with more than one degree of freedom~\cite{Schroeder}. For one quantum in two or more dimensions, two different parts of the spatial wavefunction can be entangled with each other, resulting in spatial amplitude interference, as in Young's double-slit experiment. For two quanta in one or more dimensions, the spatial wavefunctions of the two quanta can be entangled with each other, resulting in spatial intensity interference, as in the Hanbury Brown--Twiss effect~\cite{HBT1,HBT2,HBT3}. This paper will consider only the latter type of entanglement.

{Finally, these results may have potential applications in parity-time symmetry quantum control devices~\cite{Zhang}.
\section{the Gedankenexperimental setup}
Figure~\ref{fig1} shows a (1 + 1)-dimensional spacetime diagram of the Gedankenexperimental setup. In this paper ``spacetime'' always means Galilean spacetime~\cite{Penrose}. This is a lower-dimensional version of the Hanbury Brown--Twiss experiment~\cite{HBT1,HBT2,HBT3}, allowing direct visualization of the two-quanta wavefunctions and transition amplitude densities in \mbox{(2 + 1)}-dimensional configuration spacetime. {If the two detectors are both moved to the outside of the two sources, the experimental topology becomes equivalent to that of the Einstein--Podolsky--Rosen experiment~\cite{EPR}.} Configuration spacetime is the usual quantum configuration space with a time axis added. The two sources $S_a$ and $S_b$ are at fixed locations $x_a$ and $x_b$, and can each emit a single quantum on command. Let us assume we always know when a quantum is emitted. The two detectors $D_c$ and $D_d$ are at variable locations $x_c$ and $x_d$, and can each either absorb quanta or let them pass through undisturbed.

For the one-quantum cases, a single run of the Gedankenexperiment will consist of source $S_a$ emitting a single quantum at the initial time $t_i$, then this quantum either passing through both detectors, or being absorbed by one detector, or being absorbed by the other detector. Let us assume the single quantum is produced by spontaneous emission and absorbed by the time-reverse of spontaneous emission. We will do many runs, but analyze only the subset of runs where the detector $D_c$ absorbs the quantum at the final time $t_f$. There will only be one or no quanta in the apparatus at any time. The probability for all other experimental results is then one minus the probability that we will calculate.

For the two-quanta cases, a single run of the Gedankenexperiment will consist of each source emitting a single quantum, then the two quanta either passing through both detectors, or only one quantum being absorbed by one detector, or both quanta being absorbed by one detector, or one quantum being absorbed by one detector and the other quantum being absorbed by the other detector. We will do many runs, but analyze only the subset of runs where the sources each emit one quantum at the same initial time $t_i$, and the detectors each absorb one quantum at the same final time $t_f$. There will only be two or fewer quanta in the apparatus at any time. The probability for all other experimental results is then one minus the probability that we will calculate.

{This Gedankenexperiment might be experimentally realized using existing single-photon sources and semitransparent single-photon detectors.}

\begin{figure}[]
\includegraphics[width=5 cm]{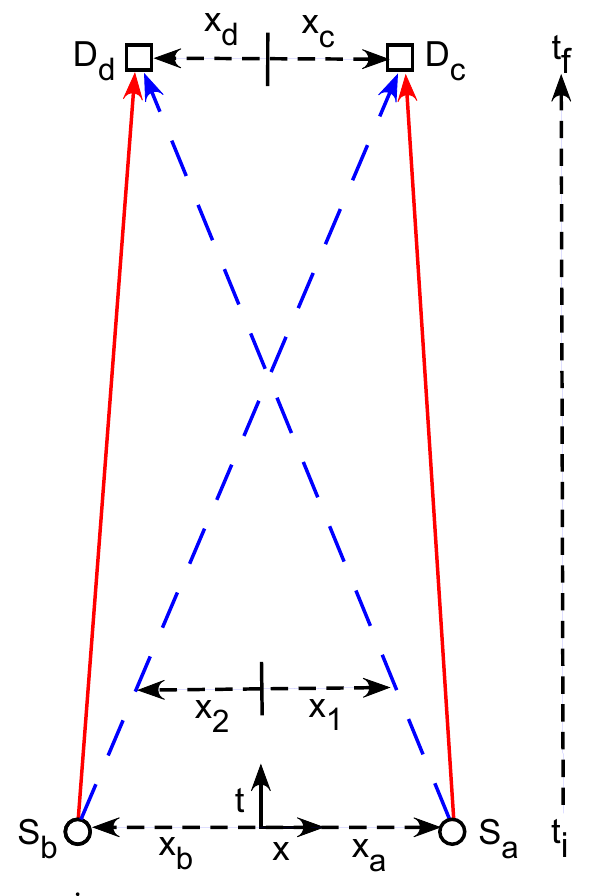}
\caption{A (1 + 1)-dimensional spacetime diagram of the Gedankenexperimental setup. The sources $S_a$ and $S_b$ are at fixed locations $x_a$ and $x_b$, while the detectors $D_c$ and $D_d$ are at variable locations $x_c$ and $x_d$. The sources and detectors are colinear in space. The sources each emit one quantum on command at the initial time $t_i$. The locations of these quanta are $x_1$ and $x_2$. We analyze only those runs where the detectors each absorb one quantum at the final time $t_f$. The red (solid) and blue (dashed) lines between the sources and detectors show two possible ways this can happen. This is a lower dimensional version of the Hanbury Brown--Twiss effect experiment~\cite{HBT1,HBT2,HBT3}. {If the two detectors are both moved to the outside of the two sources, the experimental topology becomes equivalent to that of the Einstein--Podolsky--Rosen experiment~\cite{EPR}.}\label{fig1}}
\end{figure}

\section{One Quantum}
The conventional formulation assumes that a wavefunction which lives in configuration space and evolves in time gives the most complete description of a quantum that is in principle possible. Let us assume that a quantum is created in source $S_a$ as a normalized Gaussian wavefunction $\psi$ of initial width $\sigma=1$:
\begin{equation}
\psi(x_1,t;x_a,t_i)\equiv\left(\frac{2}{\pi}\right)^{1/4}\left(\frac{1}{i(t-t_i)+2}\right)^{1/2}exp\left[-\frac{(x_1-x_a)^2}{2i(t-t_i)+4}\right],
\label{eq.1}
\end{equation}
where $x_1$ is the location of the quantum, $(x_a,t_i)=(10,0)$ are the emission location and time, all quantum masses are set to $1$, and natural units are used: $\hbar=c=1$. 

The conventional formulation assumes that upon measurement by a detector a wavefunction abruptly collapses onto a different wavefunction localized at the detector. Let us assume that at $(x_c,t_f)=(7,60)$, the wavefunction $\psi$ collapses onto the normalized gaussian wavefunction $\phi$ of width $\sigma=1$:
\begin{equation}
\phi(x_1,t;x_c,t_f)\equiv\left(\frac{2}{\pi}\right)^{1/4}\left(\frac{1}{i(t-t_f)+2}\right)^{1/2}exp\left[-\frac{(x_1-x_c)^2}{2i(t-t_f)+4}\right],
\label{eq.2}
\end{equation}
and is absorbed by detector $D_c$. I chose the fixed locations of all sources and detectors in this paper to show the symmetry of the complex transition amplitude density, to give about the same values of $P_c$ for bosons and fermions, and to give a relatively large value for $P_c$. Figure~\ref{fig2}a shows the real parts of $\psi$ and $\phi$ during the run. The imaginary parts are not shown because they do not contribute much more of interest. The conventional formulation assumes the probability for this transition is $P_c=A_c^\ast A_c$, where the subscript $c$ denotes the conventional formulation, and the conventional amplitude $A_c$ for the transition~is:
\begin{equation}
A_c=\int_{-\infty}^{\infty}\phi^\ast(x_1,60;x_c,t_f)\psi(x_1,60;x_a,t_i)dx_1,
\label{eq.3}
\end{equation}
where the time $t=60$ is the time of wavefunction collapse. Plugging in numbers gives a transition probability $P_c=6.59\times10^{-2}$ for this particular choice of source and \mbox{detector~locations.}

\nointerlineskip
\begin{figure}	
\includegraphics[width=15 cm]{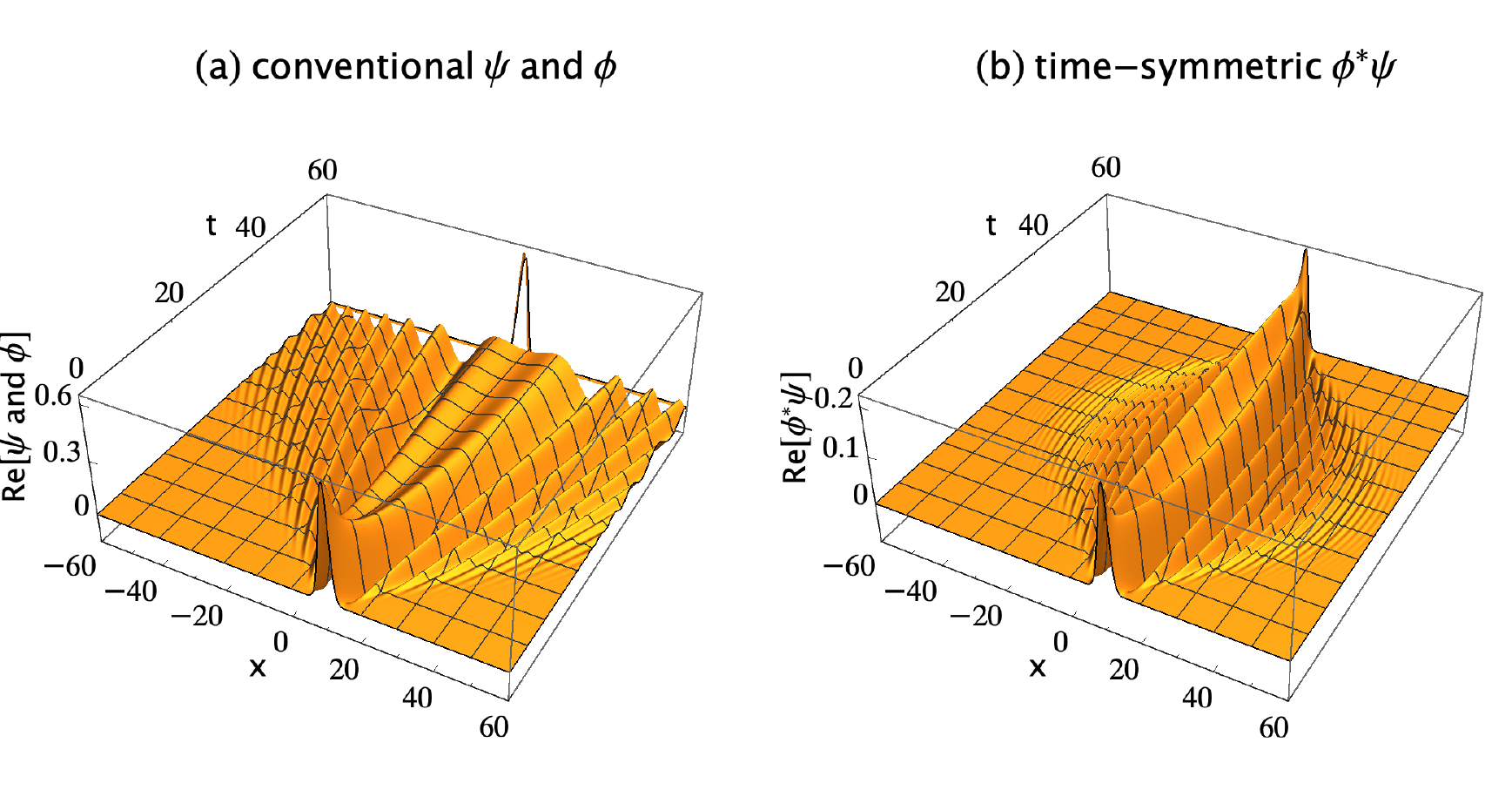}
\caption{(\textbf{a}) The conventional explanation of a Gedankenexperiment with one quantum: the one-quantum wavefunction $\psi$ is emitted by source $S_a$ at $(x_a,t_i)=(10,0)$, evolves in time, then abruptly collapses onto the wavefunction $\phi$ and is absorbed by detector $D_c$ at $(x_c,t_f)=(7,60)$. The conventional formulation assumes the wavefunction is a 1-dimensional object which lives in configuration space, evolves in time, and gives the most complete description of the quantum that is in principle possible. (\textbf{b}) The time-symmetric explanation of the same Gedankenexperiment: the one-quantum complex transition amplitude density $\phi^\ast\psi$ (where $\phi^\ast$ is the complex conjugate of the $\phi$ in the conventional explanation) is emitted by source $S_a$ and absorbed by detector $D_c$. There is no abrupt collapse. The time-symmetric formulation assumes the complex transition amplitude density is a (1 + 1)-dimensional object which lives in configuration spacetime and gives the most complete description of the quantum that is in principle possible. Configuration spacetime is the usual quantum configuration space with a time axis added. The transition amplitude density $\phi^\ast\psi$ is normalized to give a transition probability of one, and only the real parts of $\psi$, $\phi$, and $\phi^\ast\psi$ are shown.\label{fig2}}
\end{figure}

The collapse of the wavefunction at $t_f=60$ must be instantaneous, to prevent the possibility of the particle being detected in two different locations simultaneously. This instantaneous collapse violates the principle of relativistic local causality. This is the quantum nonlocality problem. One part of the quantum measurement problem is how (or whether) wavefunction collapse occurs. Another part is why the wavefunction collapses stochastically at one location and not at a different location.

The time-symmetric formulation assumes that a complex transition amplitude density which lives in configuration spacetime gives the most complete description of a quantum that is in principle possible. Using the same initial and final wavefunctions given above, the transition amplitude density $\phi^\ast\psi$ for the same transition is defined as:

\begin{equation}
\phi^\ast\psi(x_1,t;x_c,x_a,t_f,t_i)\equiv\phi^\ast(x_1,t;x_c,t_f)\psi(x_1,t;x_a,t_i),
\label{eq.4}
\end{equation}
where $\phi^\ast\psi$ varies continuously and smoothly, with no abrupt collapse, between emission at the source and absorption at the detector. The transition amplitude is the quantum amplitude for a particular transition between an initial condition and a final condition. The transition amplitude density is the quantity that is integrated over a spatial volume of configuration spacetime to get the transition amplitude. Figure~\ref{fig2}b shows the real part of $\phi^\ast\psi$ for this transition, where the probability for the transition is normalized to one. The time-symmetric formulation assumes the probability for the transition is $P_t=A_t^\ast A_t$, where the subscript $t$ denotes the time-symmetric formulation, and the time-symmetric amplitude $A_t$ is given by:
\begin{equation}
A_t=\int_{-\infty}^{\infty}\phi^\ast\psi(x_1,t;x_c,x_a,t_f,t_i)dx_1.
\label{eq.5}
\end{equation}

This has the same integrand as the conventional formulation Equation (\ref{eq.3}), except the time $t$ is now a variable. Plugging in numbers gives a transition probability $P_t=6.59\times10^{-2}$ for this particular choice of source and detector locations, the same predicted experimental result as the conventional formulation. The results are the same because the integral is independent of time. This implies the time-symmetric formulation has more time symmetry than the conventional formulation.

The transition amplitude density diverges from the source and converges to the detector, with no instantaneous collapse. This is consistent with the principle of relativistic local causality. This solves the quantum nonlocality problem and one part of the quantum measurement problem.
\section{Two Distinguishable Quanta}
Let us assume two distinguishable quanta (quanta 1 and 2) are emitted simultaneously from sources $S_a$ and $S_b$, with quantum 1 having the same initial wavefunction as in the prior one-quantum case, while quantum 2 has the similar normalized initial wavefunction:
\begin{equation}
\psi(x_2,t;x_b,t_i)\equiv\left(\frac{2}{\pi}\right)^{1/4}\left(\frac{1}{i(t-t_i)+2}\right)^{1/2}exp\left[-\frac{(x_2-x_b)^2}{2i(t-t_i)+4}\right],
\label{ }
\end{equation}where $x_2$ is the location of quantum 2, and $(x_b,t_i)=(-10,0)$ are the emission location and time at source $S_b$. We will also assume that the two initial wavefunctions abruptly collapse onto two normalized final wavefunctions (similar to the final wavefunction in the prior one-quantum case) and are absorbed by the two detectors at $(x_c,t_f)=(7,60)$ and $(x_d,t_f)=(-7,60)$. There are four possible distinguishable path permutations: (1) quantum 1 goes from $S_a$ to $D_c$, while concurrently quantum 2 goes from $S_b$ to $D_d$; (2) quantum 2 goes from $S_a$ to $D_c$, while concurrently quantum 1 goes from $S_b$ to $D_d$; (3) quantum 1 goes from $S_a$ to $D_d$, while concurrently quantum 2 goes from $S_b$ to $D_c$; and (4) quantum 2 goes from $S_a$ to $D_d$, while concurrently quantum 1 goes from $S_b$ to $D_c$. The probability for all other experimental results is then one minus the probability of these four runs. 

The conventional formulation assumes the two-quanta wavefunctions are the products of the two one-quantum wavefunctions. For the first path permutation, the two-quanta initial wavefunction is then:
\begin{equation}
\psi(x_1,x_2,t;x_a,x_b,t_i)\equiv\psi(x_1,t;x_a,t_i)\psi(x_2,t;x_b,t_i),
\label{ }
\end{equation}where $\psi(x,t;x_i,t_i)$ is defined by Equation (\ref{eq.1}), and the two-quanta collapsed \mbox{wavefunction~is}:
\begin{equation}
\phi(x_1,x_2,t;x_c,x_d,t_f)\equiv\phi(x_1,t;x_c,t_f)\phi(x_2,t;x_d,t_f),
\label{ }
\end{equation}where $\phi(x,t;x_f,t_f)$ is defined by Equation (\ref{eq.2}). Figure~\ref{fig3}a shows the real parts of the conventional initial and collapsed two-quanta wavefunctions for the first distinguishable path permutation. The imaginary parts are not shown because they do not contribute much more of interest. 

\nointerlineskip
\begin{figure}	
\includegraphics[width=15 cm]{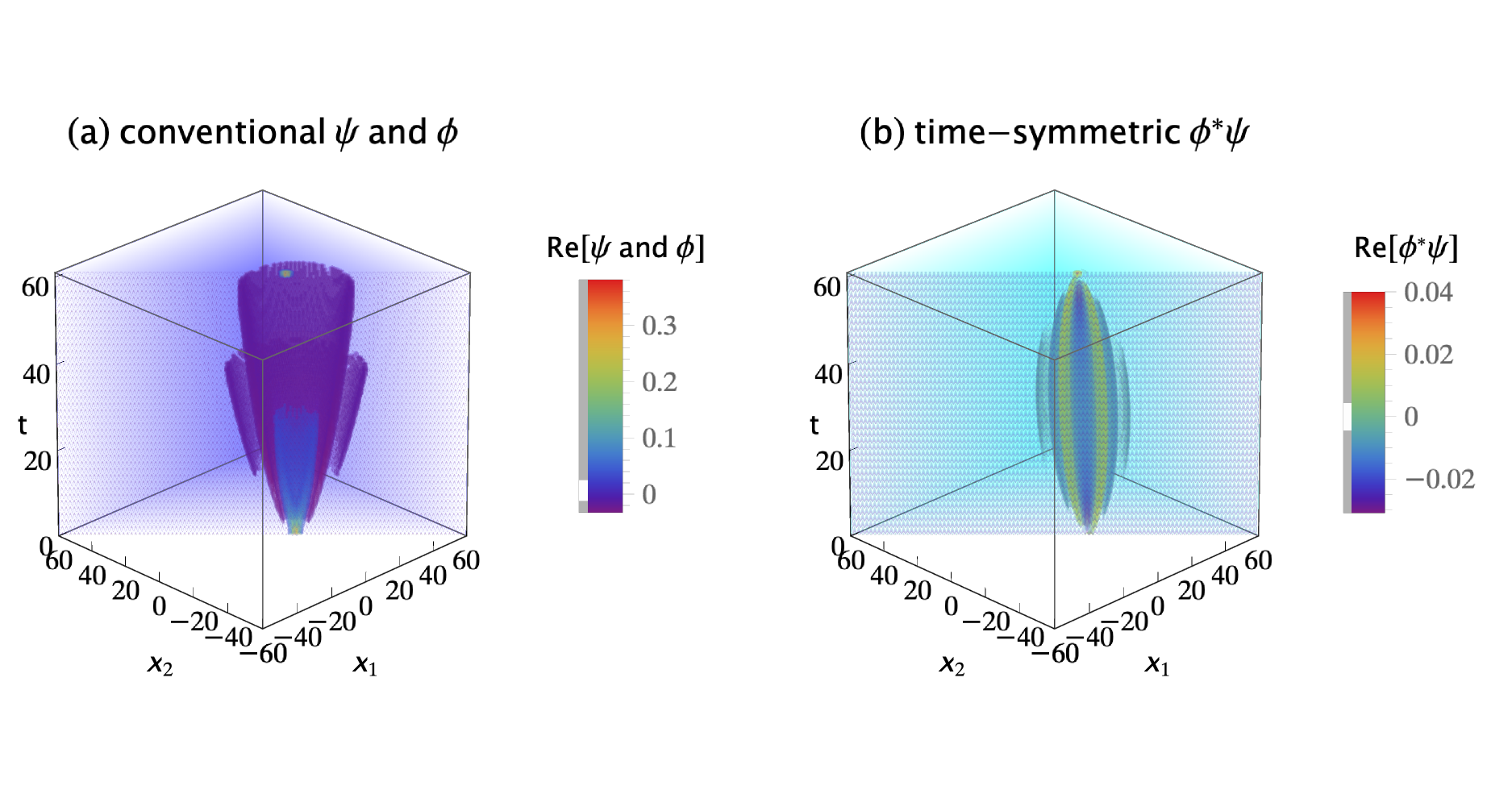}
\caption{(\textbf{a}) The conventional explanation of a Gedankenexperiment with two distinguishable quanta: only the first of the four possible distinguishable path permutations is shown. The two-quanta wavefunction $ \psi$ is emitted by sources $S_a$ at $(x_a,t_i)=(10,0)$ and $S_b$ at $(x_b,t_i)=(-10,0)$, evolves in time, then abruptly collapses onto the two-quanta wavefunction $\phi$ and is absorbed by detectors $D_c$ at $(x_c,t_f)=(7,60)$ and $D_d$ at $(x_d,t_f)=(-7,60)$. The conventional formulation assumes the two-quanta wavefunction is a 2-dimensional object which lives in configuration space, evolves in time, and gives the most complete description of the two quanta that is in principle possible. (\textbf{b}) The time-symmetric explanation of the same Gedankenexperiment: only the first of the four possible distinguishable path permutations is shown. The two-quanta transition amplitude density $\phi^\ast\psi$ (where $\phi^\ast$ is the complex conjugate of $\phi$ in the conventional explanation) is emitted by sources $S_a$ and $S_b$ and absorbed by detectors $D_c$ and $D_d$. There is no abrupt collapse. The time-symmetric formulation assumes the complex transition amplitude density is a (2 + 1)-dimensional object which lives in configuration spacetime and gives the most complete description of the two quanta that is in principle possible. The transition amplitude density $\phi^\ast\psi$ is normalized to give a transition probability of one, only the real parts of $\psi$, $\phi$, and $\phi^\ast\psi$ are shown, and half of the plots are cut away to show the interiors.\label{fig3}}
\end{figure}

The conventional formulation assumes the probability for the first distinguishable path permutation is $P_{c1}=A_{c1}^\ast A_{c1}$, where the conventional amplitude $A_{c1}$~is:
\begin{equation}
A_{c1}=\iint_{-\infty}^{\infty}\phi^\ast(x_1,x_2,60;x_c,x_d,t_f)\psi(x_1,x_2,60;x_a,x_b,t_i)dx_1dx_2,
\label{eq.9}
\end{equation}where $t=60$ is the time of wavefunction collapse. The conventional formulation assumes the total probability for any of these four events to happen is given by calculating the probability for each of these amplitudes and then adding these probabilities:
\begin{equation}
P_c=A_{c1}^\ast A_{c1}+A_{c2}^\ast A_{c2}+A_{c3}^\ast A_{c3}+A_{c4}^\ast A_{c4}.
\label{ }
\end{equation}Plugging in numbers gives $P_c=1.33\times10^{-2}$ for this particular choice of source and detector~locations.

The time-symmetric formulation assumes the two-quanta transition amplitude densities are the products of the two one-quantum transition amplitude densities. For the first distinguishable path permutation, the two-quanta transition amplitude density $\phi^\ast\psi$ is defined as:
\begin{equation}
\phi^\ast\psi(x_1,x_2,t;x_c,x_a,x_d,x_b,t_f,t_i)\equiv\phi^\ast(x_1,t;x_c,t_f)\psi(x_1,t;x_a,t_i)\phi^\ast(x_2,t;x_d,t_f)\psi(x_2,t;x_b,t_i).
\label{ }
\end{equation}

Figure~\ref{fig3}b shows the time-symmetric two-quanta transition amplitude density for the first distinguishable path permutation. Note that $\phi^\ast\psi$ varies continuously and smoothly, with no abrupt collapse, between emission at the sources and absorption at the detectors.

The time-symmetric formulation assumes the probability for the first distinguishable path permutation is $P_{t1}=A_{t1}^\ast A_{t1}$, where the time-symmetric amplitude $A_{t1}$ is given by the integral of the two-quanta transition amplitude density:
\begin{equation}
A_{t1}=\iint_{-\infty}^{\infty}\phi^\ast\psi(x_1,x_2,t;x_c,x_a,x_d,x_b,t_f,t_i)dx_1dx_2.
\label{ }
\end{equation}

This has the same integrand as the conventional formulation Equation (\ref{eq.9}), except the time $t$ is now a variable. Plugging in numbers gives $P_c=1.33\times10^{-2}$ for this particular choice of source and detector locations, the same predicted experimental result as the conventional formulation. The results are the same because the integral is independent of time. This implies the time-symmetric formulation has more time symmetry than the conventional formulation.
Figure~\ref{fig4} shows the conventional and time-symmetric predictions for how the experimentally measurable probability of a two-quanta transition will vary as a function of the positions of the two detectors, for all four distinguishable path permutations. The conventional and time-symmetric predictions are the same. There is no two-quanta interference pattern since the two quanta are distinguishable and not entangled.

\nointerlineskip
\begin{figure}	
\includegraphics[width=15 cm]{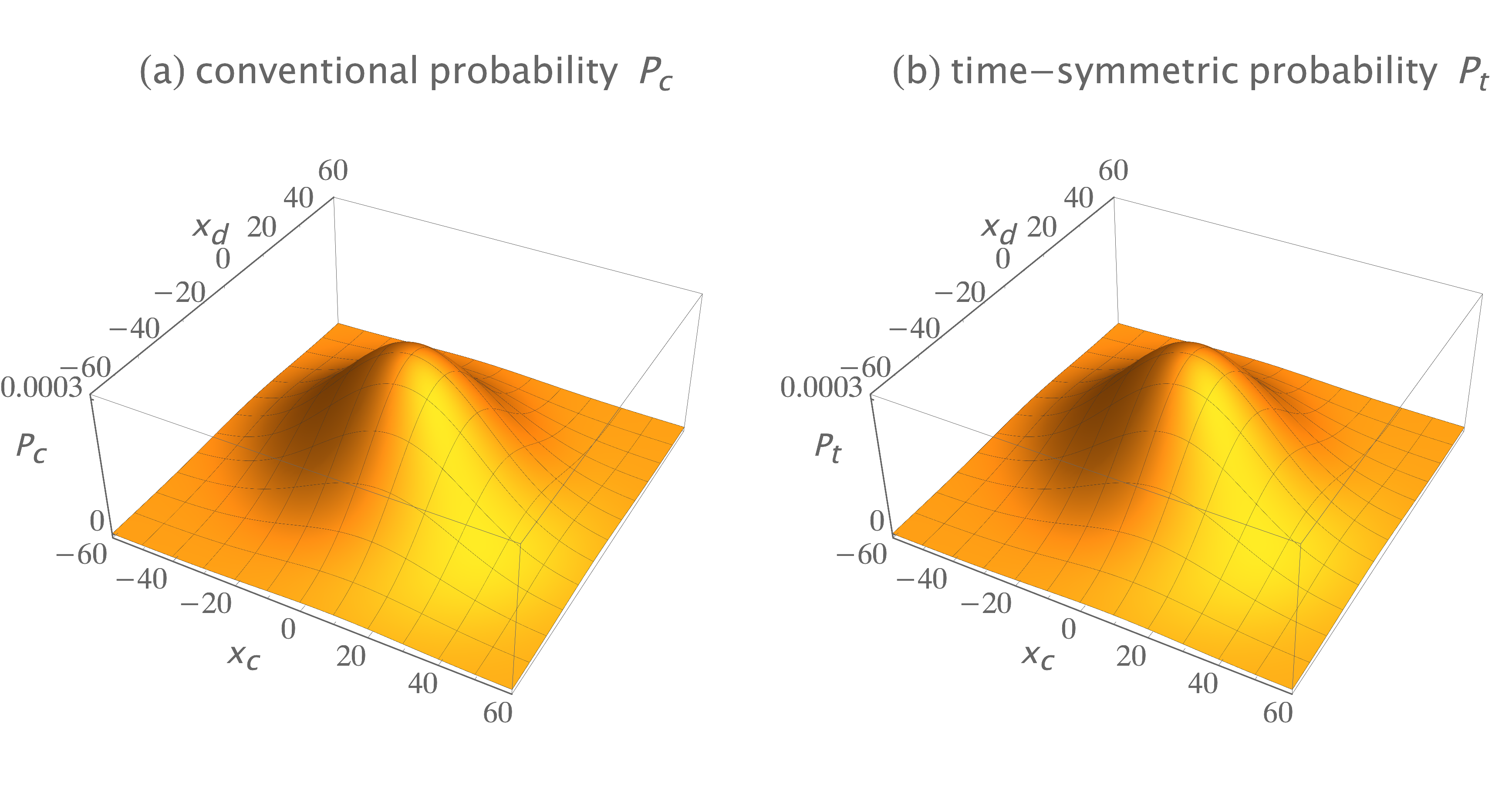}
\caption{(\textbf{a}) The conventional prediction for the Gedankenexperiment with two distinguishable quanta for all four possible distinguishable results: the probability $P_c$ that the two quanta emitted from the sources are absorbed in the two detectors as the locations of the two detectors are varied, averaged over many runs. Since the two quanta are distinguishable, there is no interference. (\textbf{b}) The time-symmetric prediction for the same Gedankenexperiment. The predictions are identical.\label{fig4}}
\end{figure}  

\section{Two Indistinguishable Bosons}\label{sec:5}
Let us assume two indistinguishable, noninteracting bosons (bosons 1 and 2) are emitted simultaneously from sources $S_a$ and $S_b$, with the same initial and collapsed two-quanta wavefunctions as in the prior distinguishable two-quanta case.

The conventional formulation assumes these two-quanta wavefunctions must be symmetrized by quantum exchange and added if they are indistinguishable bosons. The symmetrized and normalized initial two-quanta wavefunction is:


\begin{equation}
\psi_s(x_1,x_2,t;x_a,x_b,t_i)=[\psi(x_1,t;x_a,t_i)\psi(x_2,t;x_b,t_i)+\psi(x_2,t;x_a,t_i)\psi(x_1,t;x_b,t_i)]/\sqrt{2},
\label{ }
\end{equation}
where the subscript $s$ denotes symmetrization. The symmetrized and normalized two-quanta collapsed wavefunction is:
\begin{equation}
\phi_s(x_1,x_2,t;x_c,x_d,t_f)=[\phi(x_1,t;x_c,t_f)\phi(x_2,t;x_d,t_f) +\phi(x_2,t;x_c,t_f)\phi(x_1,t;x_d,t_f)]/\sqrt{2}.
\label{ }
\end{equation}

Figure~\ref{fig5}a shows the real parts of the symmetrized initial wavefunction \linebreak
$\psi_s(x_1,x_2,t;x_a,x_b,t_i)$ and the symmetrized collapsed wavefunction $\phi_s(x_1,x_2,60;x_c,x_d,t_f)$. The imaginary parts are not shown because they do not contribute much more of interest. The conventional formulation assumes the amplitude for this transition is the \mbox{overlap~integral:}
\begin{equation}
A_{c}=\iint_{-\infty}^{\infty}\phi^\ast_s(x_1,x_2,60;x_c,x_d,t_f)\psi_s(x_1,x_2,60;x_a,x_b,t_i)dx_1dx_2.
\label{ }
\end{equation}where $t=60$ is the time of wavefunction collapse. The conventional probability is \linebreak
 $P_c=A^\ast_{c}A_{c}$. Plugging in numbers gives $P_c=6.25\times10^{-3}$ for this particular choice of source and detector locations.


\nointerlineskip
\begin{figure}	

\includegraphics[width=15 cm]{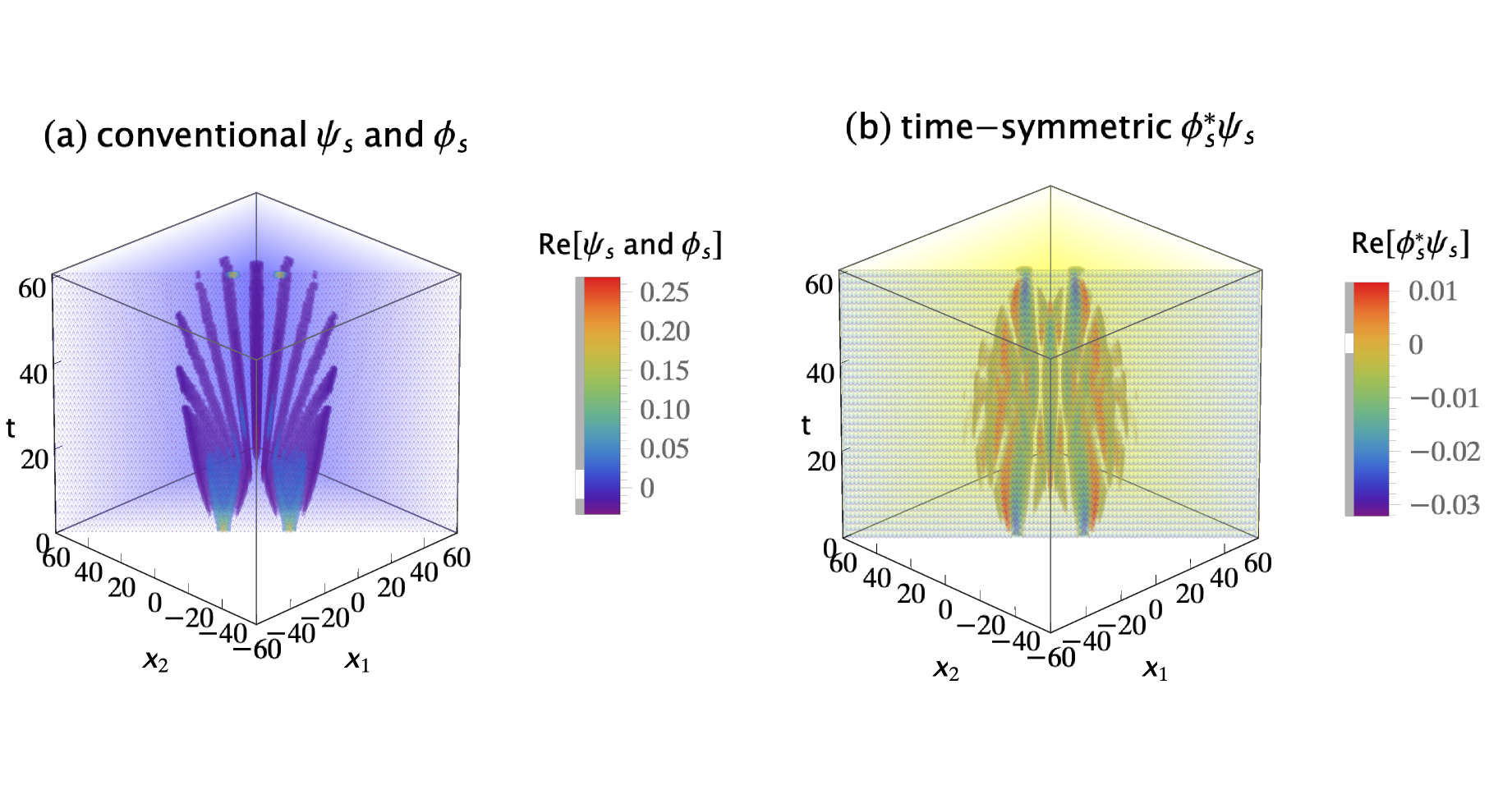}
\caption{(\textbf{a}) The conventional explanation of a Gedankenexperiment with two indistinguishable bosons: the symmetrized two-quanta wavefunction $\psi_s$ is emitted by sources $S_a$ at $(x_a,t_i)=(10,0)$ and $S_b$ at $(x_b,t_i)=(-10,0)$, evolves in time, then abruptly collapses onto the symmetrized two-quanta wavefunction $\phi_s$ and is absorbed by detectors $D_c$ at $(x_c,t_f)=(7,60)$ and $D_d$ at $(x_d,t_f)=(-7,60)$. The conventional formulation assumes the two-quanta wavefunction is a 2-dimensional object which lives in configuration space, evolves in time, and gives the most complete description of the two quanta that is in principle possible. (\textbf{b}) The time-symmetric explanation of the same Gedankenexperiment: the symmetrized two-quanta transition amplitude density $\phi^\ast_s\psi_s$ (where $\phi^\ast_s$ is the complex conjugate of the $\phi_s$ in the conventional explanation) is emitted by sources $S_a$ and $S_b$, and the quanta are absorbed by detectors $D_c$ and $D_d$. There is no abrupt collapse. The time-symmetric formulation assumes the symmetrized complex transition amplitude density is a (2 + 1)-dimensional object which lives in configuration spacetime and gives the most complete description of the two quanta that is in principle possible. The transition amplitude density $\phi^\ast\psi$ is normalized to give a transition probability of one, only the real parts of $\psi$, $\phi$, and $\phi^\ast\psi$ are shown, and half of the plots are cut away to show the interiors.\label{fig5}}
\end{figure}  


The time-symmetric formulation assumes the two-quanta transition amplitude densities must be symmetrized by path exchange and added if they are indistinguishable bosons. There are four possible indistinguishable path permutations: (1) quantum 1 goes from $S_a$ to $D_c$, while concurrently quantum 2 goes from $S_b$ to $D_d$; (2) quantum 2 goes from $S_a$ to $D_c$, while concurrently quantum 1 goes from $S_b$ to $D_d$; (3) quantum 1 goes from $S_a$ to $D_d$, while concurrently quantum 2 goes from $S_b$ to $D_c$; and (4) quantum 2 goes from $S_a$ to $D_d$, while concurrently quantum 1 goes from $S_b$ to $D_c$. The sign of each permutation is positive for bosons, giving a symmetrized and normalized transition amplitude density of:

\begin{gather}
\phi^\ast_s\psi_s(x_1,x_2,t;x_c,x_a,x_d,x_b,t_f,t_i)=\nonumber \\ 
[\phi^\ast(x_1,t;x_c,t_f)\psi(x_1,t;x_a,t_i)\phi^\ast(x_2,t;x_d,t_f)\psi(x_2,t;x_b,t_i)\nonumber \\
+\thickspace\phi^\ast(x_2,t;x_c,t_f)\psi(x_2,t;x_a,t_i)\phi^\ast(x_1,t;x_d,t_f)\psi(x_1,t;x_b,t_i) \nonumber \\
+\thickspace\phi^\ast(x_1,t;x_d,t_f)\psi(x_1,t;x_a,t_i)\phi^\ast(x_2,t;x_c,t_f)\psi(x_2,t;x_b,t_i)\nonumber \\
\quad\thickspace\thinspace+\thickspace\phi^\ast(x_2,t;x_d,t_f)\psi(x_2,t;x_a,t_i)\phi^\ast(x_1,t;x_c,t_f)\psi(x_1,t;x_b,t_i)]/2.
\label{ }
\end{gather}

Note that the time-symmetric normalization constant is $1/2$, because there are four terms. Figure~\ref{fig5}b shows the symmetrized transition amplitude density $\phi^\ast_s\psi_s$. It varies continuously and smoothly, with no abrupt collapse, between emission at the sources and absorption at the detectors. The time-symmetric formulation assumes the probability of the transition is $P_t = A^\ast_t A_t$, where the amplitude $A_t$ is the integral:
\begin{equation}
A_t=\iint_{-\infty}^{\infty}\phi^\ast_s\psi_s(x_1,x_2,t;x_c,x_a,x_d,x_b,t_f,t_i)dx_1dx_2.
\label{ }
\end{equation}

This has the same integrand as the conventional formulation Equation (15), except the time $t$ is now a variable. Plugging in numbers gives $P_t=6.25\times10^{-3}$ for this particular choice of source and detector locations, the same predicted experimental result as the conventional formulation. The results are the same because the integral is independent of time. This implies the time-symmetric formulation has more time symmetry than the conventional formulation.

Figure~\ref{fig6} shows the conventional and time-symmetric predictions for how the experimentally measurable probability of a two-quanta transition will vary as a function of the positions of the two detectors. The conventional and time-symmetric predictions are the same. The two-quanta interference pattern has a maximum when the two detectors are located at $(x_c,x_d)=(0,0)$, as expected for indistinguishable bosons.

\nointerlineskip
\begin{figure}	

\includegraphics[width=15 cm]{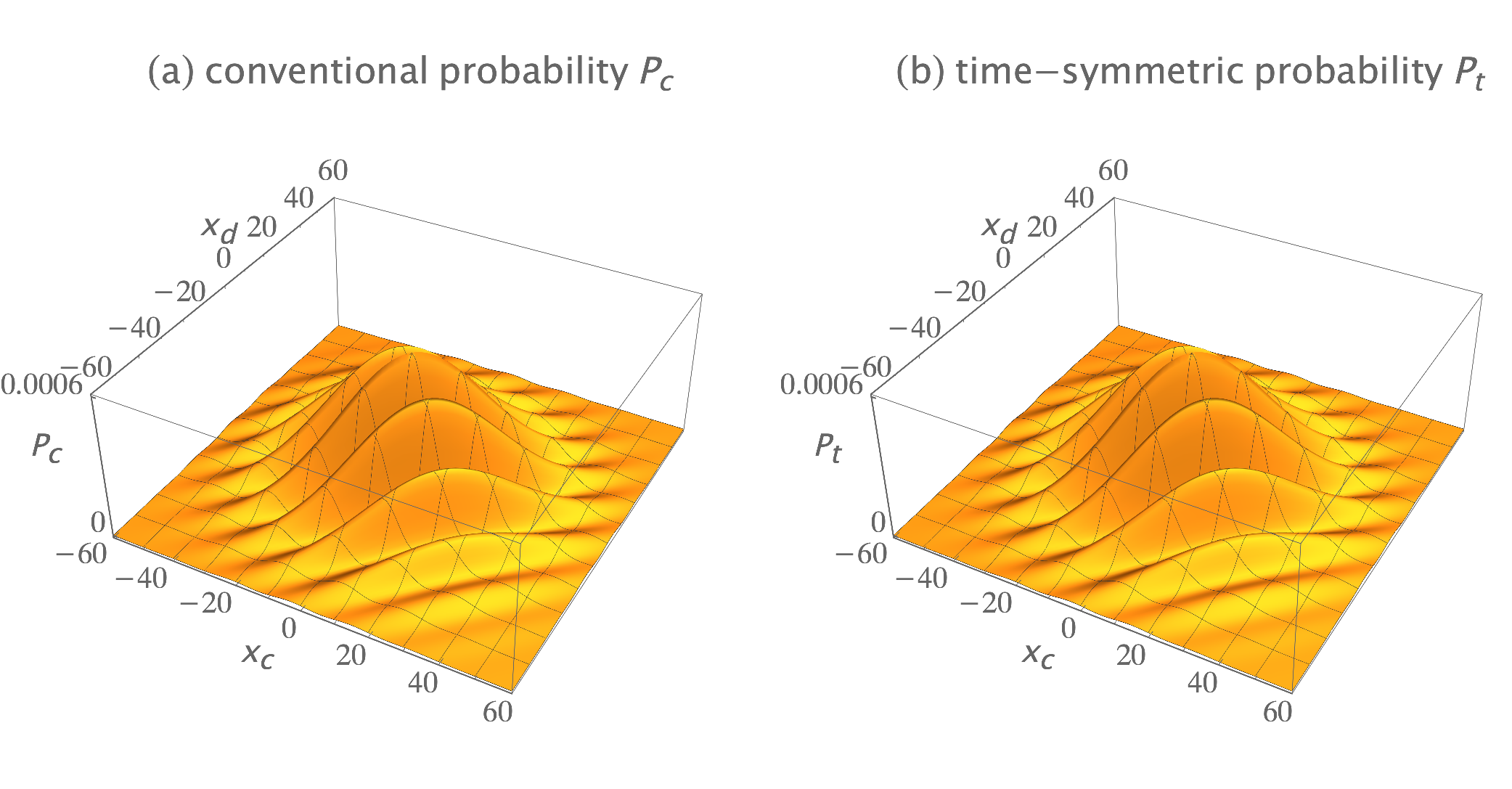}
\caption{(\textbf{a}) The conventional formulation prediction for the interference pattern for two indistinguishable bosons: the probability $P_c$ that the two quanta emitted from the sources are absorbed in the two detectors, as the locations of the two detectors are varied, averaged over many runs. (\textbf{b}) The time-symmetric formulation prediction for the same Gedankenexperiment: the probability $P_t$ that the two quanta emitted from the sources are absorbed in the two detectors, as the locations of the two detectors are varied, averaged over many runs. The interference patterns are identical, and have a maximum when the two detectors are located at $(x_c,x_d)=(0,0)$, as expected for indistinguishable bosons. The interference patterns are normalized to give a transition probability of one.\label{fig6}}
\end{figure}  


\section{Two Indistinguishable Fermions}
Let us assume two indistinguishable, noninteracting fermions (fermions 1 and 2) are emitted simultaneously from sources $S_a$ and $S_b$, with the same initial and collapsed two-quanta wavefunctions as in the earlier distinguishable two-quanta case.


The conventional formulation assumes these two-quanta wavefunctions must be antisymmetrized by quantum exchange if they are indistinguishable fermions. The antisymmetrized and normalized initial two-quanta wavefunction is:
\begin{equation}
\psi_a(x_1,x_2,t;x_a,x_b,t_i)=[\psi(x_1,t;x_a,t_i)\psi(x_2,t;x_b,t_i)-\psi(x_2,t;x_a,t_i)\psi(x_1,t;x_b,t_i)]/\sqrt{2},
\label{ }
\end{equation}
where the subscript $a$ denotes antisymmetrization. The antisymmetrized and normalized two-quanta collapsed wavefunction is:
\begin{equation}
\phi_a(x_1,x_2,t;x_c,x_d,t_f)=[\phi(x_1,t;x_c,t_f)\phi(x_2,t;x_d,t_f)-\phi(x_2,t;x_c,t_f)\phi(x_1,t;x_d,t_f)]/\sqrt{2}.
\label{ }
\end{equation}

Figure~\ref{fig7}a shows the real parts of the antisymmetrized initial wavefunction\linebreak
 $\psi_a(x_1,x_2,t;x_a,x_b,t_i)$ and the antisymmetrized collapsed wavefunction $\phi_a(x_1,x_2,t;x_c,x_d,t_f)$. The imaginary parts are not shown because they do not contribute much more of interest. The conventional formulation assumes the probability for this transition is $P_c=A^\ast_cA_c$, where the amplitude $A_c$ is given by the overlap integral:
\begin{equation}
A_c=\iint_{-\infty}^{\infty}\phi^\ast_a(x_1,x_2,60;x_c,x_d,t_f)\psi_a(x_1,x_2,60;x_a,x_b,t_i)dx_1dx_2,
\label{eq.20}
\end{equation}where the time $t=60$ is the time of wavefunction collapse. Plugging in numbers gives $P_c=7.09\times10^{-3}$ for this particular choice of source and detector locations.


\nointerlineskip
\begin{figure}	

\includegraphics[width=15 cm]{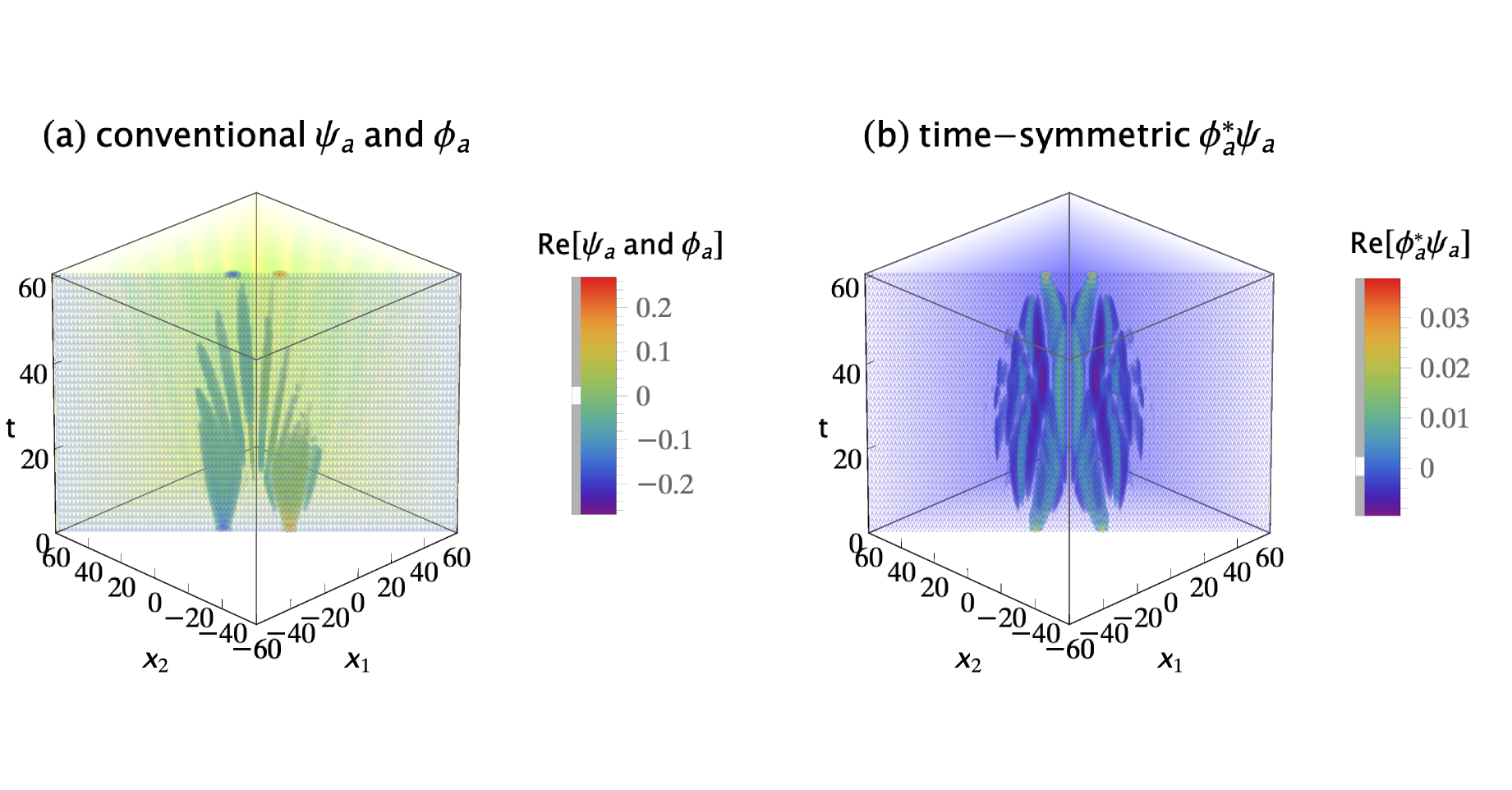}
\caption{(\textbf{a}) The conventional explanation of a Gedankenexperiment with two indistinguishable fermions: the antisymmetrized two-quanta wavefunction $\psi_a$ is emitted by sources $S_a$ at $(x_a,t_i)=(10,0)$ and $S_b$ at $(x_b,t_i)=(-10,0)$, evolves in time, then abruptly collapses onto the antisymmetrized two-quanta wavefunction $\phi_a$ and is absorbed by detectors $D_c$ at $(x_c,t_f)=(7,60)$ and $D_d$ at $(x_d,t_f)=(-7,60)$. The conventional formulation assumes the antisymmetrized wavefunction is a 2-dimensional object which lives in configuration space, evolves in time, and gives the most complete description of the two quanta that is in principle possible. (\textbf{b}) The time-symmetric explanation of the same Gedankenexperiment: the antisymmetrized two-quanta transition amplitude density $\phi^\ast_a\psi_a$ (where $\phi^\ast_a$ is the complex conjugate of the $\phi_a$ in the conventional explanation) is emitted by sources $S_a$ and $S_b$, and the quanta are absorbed by detectors $D_c$ and $D_d$. There is no abrupt collapse. The time-symmetric formulation assumes the antisymmetrized complex transition amplitude density is a (2 + 1)-dimensional object which lives in configuration spacetime and gives the most complete description of the two quanta that is in principle possible. The transition amplitude density $\phi^\ast\psi$ is normalized to give a transition probability of one, only the real parts of $\psi$, $\phi$, and $\phi^\ast\psi$ are shown, and half of the plots are cut away to show the interiors.\label{fig7}}
\end{figure}  


The time-symmetric formulation assumes the two-quanta transition amplitude densities must be antisymmetrized by path exchange if they are indistinguishable fermions. There are four possible indistinguishable path permutations: (1) quantum 1 goes from $S_a$ to $D_c$, while concurrently quantum 2 goes from $S_b$ to $D_d$; (2) quantum 2 goes from $S_a$ to $D_c$, while concurrently quantum 1 goes from $S_b$ to $D_d$; (3) quantum 1 goes from $S_a$ to $D_d$, while concurrently quantum 2 goes from $S_b$ to $D_c$; and (4) quantum 2 goes from $S_a$ to $D_d$, while concurrently quantum 1 goes from $S_b$ to $D_c$. The sign of each permutation is determined by the number of path termini pairs that are exchanged: using the first transition as the reference, the first permutation has zero termini pairs swapped; the second permutation has two termini pairs swapped; the third permutation has one termini pair swapped; and the fourth permutation has one termini pair swapped. Assigning positive signs to even termini swap permutations and negative signs to odd termini swap permutations gives an antisymmetrized and normalized transition amplitude density of:
\begin{gather}
\phi^\ast_a\psi_a(x_1,x_2,t;x_c,x_a,x_d,x_b,t_f,t_i)=\nonumber \\ 
[\phi^\ast(x_1,t;x_c,t_f)\psi(x_1,t;x_a,t_i)\phi^\ast(x_2,t;x_d,t_f)\psi(x_2,t;x_b,t_i)\nonumber \\
+\thickspace\phi^\ast(x_2,t;x_c,t_f)\psi(x_2,t;x_a,t_i)\phi^\ast(x_1,t;x_d,t_f)\psi(x_1,t;x_b,t_i) \nonumber \\
-\thickspace\phi^\ast(x_1,t;x_d,t_f)\psi(x_1,t;x_a,t_i)\phi^\ast(x_2,t;x_c,t_f)\psi(x_2,t;x_b,t_i)\nonumber \\
\quad\thickspace\thinspace-\thickspace\phi^\ast(x_2,t;x_d,t_f)\psi(x_2,t;x_a,t_i)\phi^\ast(x_1,t;x_c,t_f)\psi(x_1,t;x_b,t_i)]/2.
\label{ }
\end{gather}

Note that the time-symmetric normalization constant is $1/2$, because there are four terms. Figure~\ref{fig7}b shows the antisymmetrized transition amplitude density $\phi^\ast_a\psi_a$. It varies continuously and smoothly, with no abrupt collapse, between emission at the sources and absorption at the detectors. The time-symmetric formulation assumes the probability of the transition is $P_t = A^\ast_t A_t$, where the amplitude $A_t$ is the integral:
\begin{equation}
A_t=\iint_{-\infty}^{\infty}\phi^\ast_a\psi_a(x_1,x_2,t;x_c,x_a,x_d,x_b,t_f,t_i)dx_1dx_2.
\label{eq.22}
\end{equation}

This has the same integrand as the conventional formulation Equation (\ref{eq.20}), except the time $t$ is now a variable. Plugging in numbers gives $P_t=7.09\times10^{-3}$ for this particular choice of source and detector locations, the same predicted experimental result as the conventional formulation. The results are the same because the integral is independent of time. This implies the time-symmetric formulation has more time symmetry than the conventional formulation.

Figure~\ref{fig8} shows the conventional and time-symmetric predictions for how the experimentally measurable probability of a two-quanta transition will vary as a function of the positions of the two detectors. The conventional and time-symmetric predictions are the same. The two-quanta interference pattern has a minimum when the two detectors are located at $(x_c,x_d)=(0,0)$, as expected for indistinguishable fermions.

\nointerlineskip
\begin{figure}	

\includegraphics[width=15 cm]{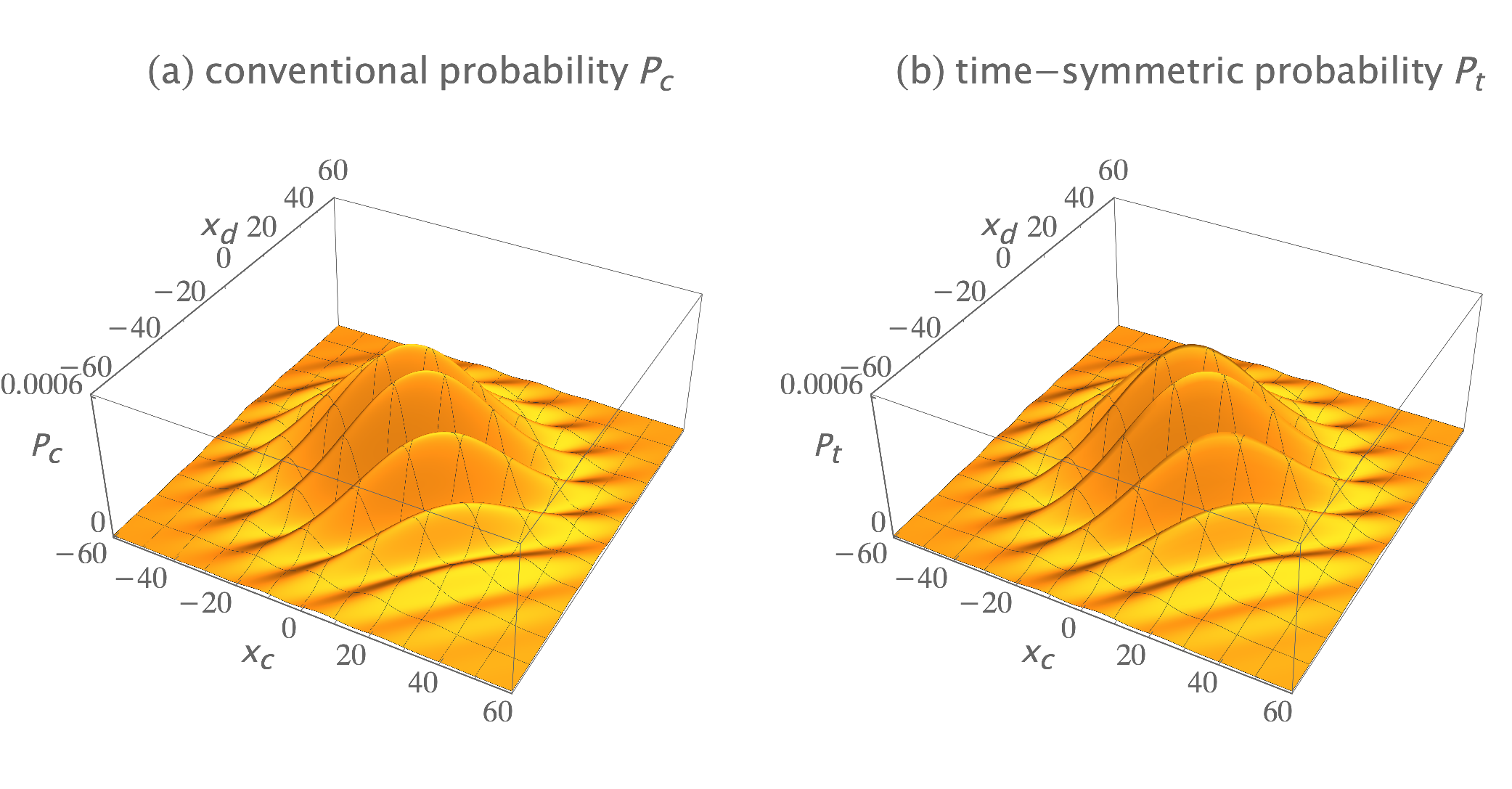}
\caption{(\textbf{a}) The conventional formulation prediction for the interference pattern for two indistinguishable bosons: the probability $P_c$ that the two quanta emitted from the sources are absorbed in the two detectors, as the locations of the two detectors are varied, averaged over many runs. (\textbf{b}) The time-symmetric formulation prediction for the same Gedankenexperiment: the probability $P_t$ that the two quanta emitted from the sources are absorbed in the two detectors, as the locations of the two detectors are varied, averaged over many runs. The interference patterns are identical, and have a maximum when the two detectors are located at $(x_c,x_d)=(0,0)$, as expected for indistinguishable bosons. The interference patterns are normalized to give a transition probability of one.\label{fig8}}
\end{figure}  


\section{The Original Quantum Analysis of the Hanbury Brown--Twiss Effect}
The Hanbury Brown--Twiss effect was initially demonstrated with radio waves and explained using classical electromagnetic theory~\cite{HBT1,HBT2,HBT3}. When Hanbury Brown and Twiss proposed using their effect to measure stellar diameters with optical photons, most physicists believed it would not work because optical photons were much more like particles than waves. In response to Feynman's telling him ``It will never work!'', Hanbury Brown replied  ``Yes, I know. We were told so. But we built it anyway, and it did work~\cite{Radhakrishnan}.''

{The Hanbury Brown--Twiss work started a new and important branch of physics: quantum optics. People discovered that light did not always obey Maxwell's equations: it became necessary to understand the quantum nature of light. This was done largely by Klauder, Sudarshan, Glauber, and Mandel in the 50's and 60's. This led to the "coherent state" theory of light which explained the differences between thermal light, laser light, squeezed light, and other types of light. Later developments led to Raman spectroscopy, optical traps, Doppler cooling, Bose--Einstein condensation, and quantum information technologies~\cite{Wikipedia}.} 

 Fano~\cite{Fano} gave the first completely quantum explanation of the Hanbury Brown--Twiss effect. Fano described two excited atoms, $a$ and $b$, each emitting one photon, followed by two ground-state atoms, $c$ and $d$, each absorbing one photon. He drew two diagrams for the indistinguishable ways this could happen: (1) photon 1 goes from $a$ to $c$, while concurrently photon 2 goes from $b$ to $d$; and (2) photon 1 goes from $a$ to $d$, while concurrently photon 2 goes from $b$ to $c$. He then added these two alternatives to get the unnormalized total amplitude $A_\Sigma$ for the transition:

\begin{gather}
A_\Sigma=[A_{ca}A_{db}+A_{da}A_{cb}],
\label{ }
\end{gather}
where $A_{ij}$ is the amplitude for a photon to go from $j$ to $i$. Feynman~\cite{Feynman} and Mandel~\cite{Mandel} later gave the same quantum explanation as Fano. In the time-symmetric formulation of this paper, this corresponds to the unnormalized transition amplitude density:

\begin{gather}
\phi^\ast_s\psi_s(x_1,x_2,t;x_c,x_a,x_d,x_b,t_f,t_i)=\nonumber \\ 
[\phi^\ast(x_1,t;x_c,t_f)\psi(x_1,t;x_a,t_i)\phi^\ast(x_2,t;x_d,t_f)\psi(x_2,t;x_b,t_i)\nonumber\\
+\phi^\ast(x_2,t;x_d,t_f)\psi(x_2,t;x_a,t_i)\phi^\ast(x_1,t;x_c,t_f)\psi(x_1,t;x_b,t_i)].
\label{eq.24}
\end{gather}

Figure~\ref{fig9}a shows a plot of the real parts of the Fano, Feynman, and Mandel transition amplitude density $\phi^\ast_s\psi_s$, normalized to give a transition probability of one. The imaginary parts are not shown because they do not contribute much more of interest. Note that the transition amplitude density has one terminus on the $t=0$ plane, at $(x_a,x_b)=(10,-10)$, and two termini on the $t=60$ plane, at $(x_a,x_b)=(7,-7)$ and $(-7,7)$. It obviously does not have the time symmetry of Figure~\ref{fig5}b, suggesting the analyses of Fano, Feynman, and Mandel are incomplete. This could also be inferred from Figure~\ref{fig5}a. In essence, they symmetrized the final state $\phi_s$ but not the initial state $\psi$.

\nointerlineskip
\begin{figure}	

\includegraphics[width=15 cm]{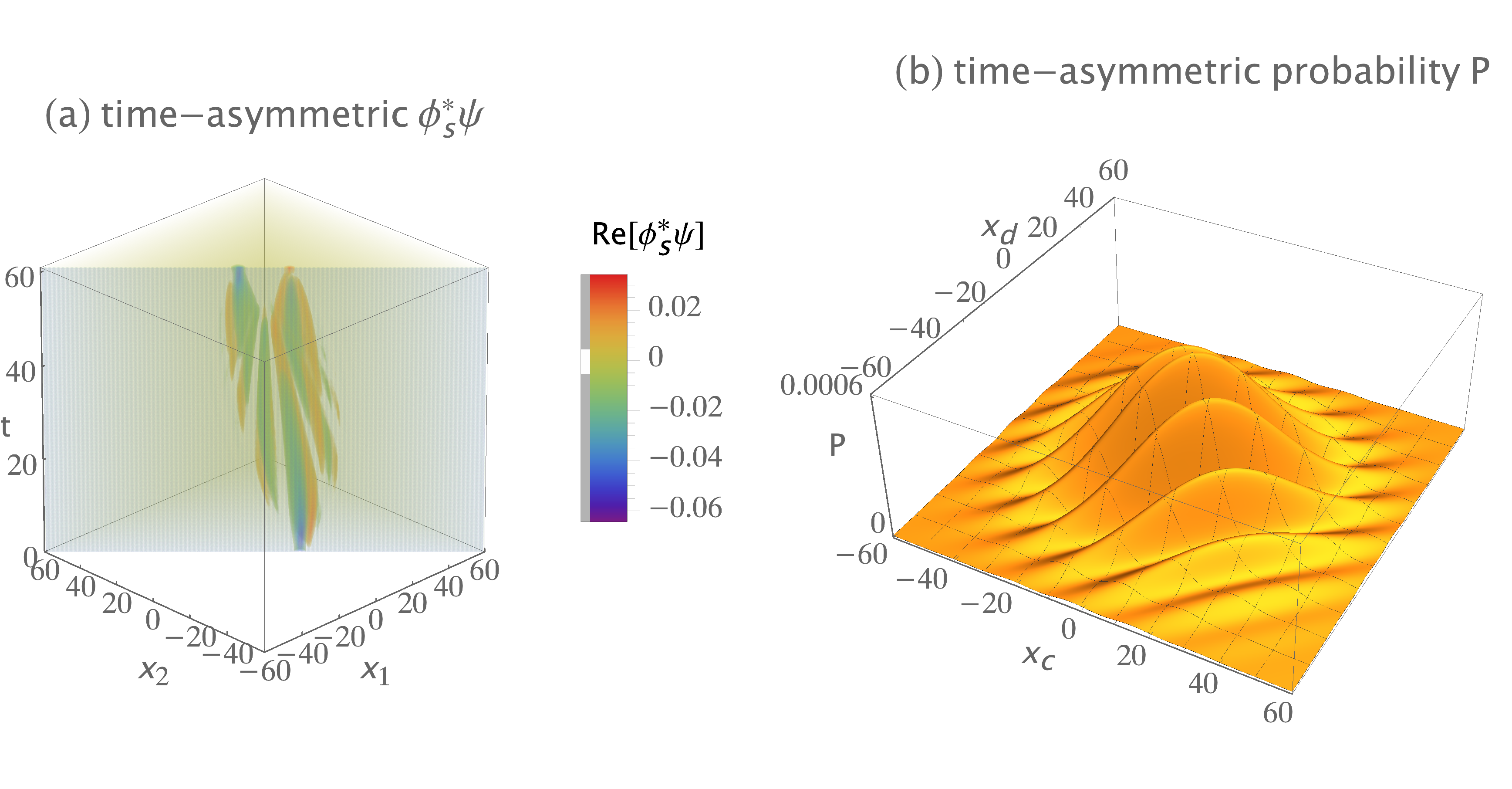}
\caption{The (1 + 1)-dimensional predictions of Fano, Feynman, and Mandel's analyses of the Hanbury Brown--Twiss experiment, for the same locations of sources and detectors used earlier. They assumed only two possible indistinguishable path permutations: quantum 1 goes from $S_a$ to $D_c$, while concurrently quantum 2 goes from $S_b$ to $D_d$; and quantum 1 goes from $S_a$ to $D_d$, while concurrently quantum 2 goes from $S_b$ to $D_c$. (\textbf{a}) The two-quanta transition amplitude density $\phi^\ast_s\psi$ is time asymmetric: compare to Figure \ref{fig5}b. This is because two other possible indistinguishable path permutations are missing. The complex transition amplitude density $\phi^\ast_s\psi$ is normalized to give a transition probability of one, only the real part of $\phi^\ast_s\psi$ is shown, and half of the $\phi^\ast_s\psi$ plot is cut away to show the interior. (\textbf{b}) The predicted probability of the transition as a function of the detector locations. It is identical to the time-symmetric experimental prediction because it is normalized to give a transition probability of one.\label{fig9}}
\end{figure}

If we normalize Equation (\ref{eq.24}) and calculate the probability for the transition, we get $P_t=3.125\times10^{-3}$, which is half the predicted experimental result of the conventional and time-symmetric formulations described in Section \ref{sec:5}. Figure~\ref{fig9}b shows their predictions for how the experimentally measurable probability of a two-quanta transition will vary as a function of the positions of the two detectors. This is identical to Figure~\ref{fig6}a,b. The differences in the predicted probability of the transition do not show up in Figure~\ref{fig9}a,b because the factors of two differences are absorbed into the normalization constants. But a comparison of experimental data with the predicted transition probabilities should distinguish between their analyses and my analysis.
\section{Discussion}
Bell asked [A.J. Leggett, private communication] ``How do you convert an 'and' into an 'or'?'' He wondered how a quantum superposition of several possible final states turns into only one final state upon measurement. The conventional formulation postulates abrupt collapse of the wavefunction upon measurement onto only one of the possible final states, while the time-symmetric formulation postulates the smoothly varying existence of only one actual transition amplitude density out of a statistical ensemble of possible transition amplitude densities, with no change in the actual one after information about it is gained from the experimental results. Note that the continuous localization of the transition amplitude density inside the detector allows the quantum to be locally absorbed by the detector, which is a well-understood process that is qualitatively different than the nonlocal wavefunction collapse of the conventional formulation, which must be instantaneous in all reference frames to obey the conservation laws. The conventional formulations ``quantum indeterminacy'' about the outcome of an experiment is replaced by the time-symmetric formulations classical uncertainty about which possible transition amplitude density actually exists. The time-symmetric formulations answer to Bell's question is that there was never an ``and,'' there was only an ``or.'' The question of how nature chooses one transition amplitude density out of a statistical ensemble is answered by the stochastic nature of the spontaneous emission and absorption processes. 

The conventional formulation has several asymmetries in time: only the initial conditions of the wavefunction are specified, the wavefunction is evolved only forward in time, the transition probability is calculated only at the time of measurement, wavefunction collapse happens only at the time of measurement, and wavefunction collapse happens only forwards in time. This seems unphysical: shouldn't the fundamental laws of nature be time-symmetric? Consider the details of a specific example: according to the conventional formulation, Equation (\ref{eq.3}) must be evaluated only at the time of the collapse. In contrast, according to the time-symmetric formulation, the transition amplitude of Equation (\ref{eq.5}) can be evaluated at any time. But the two transition amplitudes give the same results. The fact that the transition amplitude need not be evaluated at a special time shows that quantum mechanics has more intrinsic symmetry than allowed by the conventional formulation. Heisenberg said, ``Since the symmetry properties always constitute the most essential features of a theory, it is difficult to see what would be gained by omitting them in the corresponding language~\cite{Heisenberg}.'' The intrinsic time symmetry of a quantum transition is represented in the time-symmetric formulation, but not in the conventional formulation.

More generally, the conventional formulation implicitly assumes that quantum mechanics is only a predictive theory. As Dyson pointed out~\cite{Dyson}, 
statements about the past cannot in general be made in [the conventional formulation of] quantum-mechanical language. For example, we can describe a uranium nucleus by a wavefunction including an outgoing alpha particle wave which determines the probability that the nucleus will decay tomorrow. But we cannot describe by means of a wavefunction the statement, ``This nucleus decayed yesterday at 9 a.m. Greenwich time.'' Feynman also believed that the conventional formulation could not account for history~\cite{Feynman3}. When the conventional formulation is used retrodictively, attempting to determine what happened in the past given the present wavefunction, it usually does not work. Penrose~\cite{Penrose} used an interferometer Gedankenexperiment to show that using the conventional formulation retrodictively gives us ``completely the wrong answer!'' Hartle~\cite{Hartle} proved that in the conventional formulation ``correct probabilities for the past cannot generally be constructed simply by running the Schr\"{o}dinger equation backwards in time from the present state.'' This inability of the conventional formulation to describe or retrodict the past seems like a serious shortcoming for a theory that claims to be our best description of nature. Since the time-symmetric formulation is intrinsically time-symmetric, it describes the future and past equally well and makes correct predictions and retrodictions. For example, consider the single-quantum Gedankenexperiment shown in Figure~\ref{fig2}. Given the wavefunction $\psi$ at $t_i=0$ for Figure~\ref{fig2}a, the conventional formulation can correctly predict the wavefunction up until $t<60$, but not later. Given the collapsed wavefunction $\phi$ at $t\ge60$, the conventional formulation cannot retrodict the earlier wavefunction. In contrast, given the complex transition amplitude density of Figure~\ref{fig2}b at any time, the time-symmetric formulation can predict and retrodict the complex transition amplitude density at any other time.

Finally, the longstanding conceptual problems in the foundations of the conventional formulation suggest that something is fundamentally wrong. This paper proposes what might be wrong: the conventional formulation assumes the most complete description of a quantum system that is in principle possible is a wavefunction, which is an n-dimensional object which lives in configuration space and evolves only forwards in time. It is ingrained in human experience and intuition that nature is composed of 3-dimensional objects and has an intrinsic arrow of time, which leads us to implicitly extrapolate these concepts to the quantum level. These questionable extrapolations seem to be the cause of many conceptual problems in the conventional formulation. The time-symmetric formulation assumes the most complete description of a quantum system that is in principle possible is a complex transition amplitude density, which is an (n + 1)-dimensional object which lives in configuration spacetime, and does not have many of these conceptual problems.

\vfill
\end{document}